\documentclass[twocolumn,trackchanges]{aastex7}

\begin{document}

\title{No Evidence of Anomalous Diffusion in Yukawa Crystals}

\author[orcid=0009-0006-2122-5606]{M. E. Caplan}
\affiliation{Department of Physics, Illinois State University, IL, USA}
\email[show]{mecapl1@ilstu.edu}  

\author[orcid=0009-0006-3093-8317]{D. Yaacoub}
\affiliation{Department of Physics, Illinois State University, IL, USA}
\email{}

\begin{abstract}

Diffusion in Yukawa crystals is stochastic due to the thermally activated formation of vacancy-interstitial pairs, which have poor statistics in simulations. This makes it difficult to argue if Yukawa crystals exhibit normal diffusion, or if they could be subdiffusive or superdiffusive. To resolve this, we run a long molecular dynamics simulation of an idealized Yukawa crystal for a billion timesteps. We find no evidence of anomalous diffusion in the pure crystal, but also caution readers against overinterpreting this result as real crystals have complicated structures including grains and defects.

\end{abstract}



\section{Introduction} 

The elastic evolution of neutron star crusts in response to stress is in part set by microphysics including diffusion coefficients in the crystallized plasma. 
In our recent Letter, we have resolved diffusion coefficients in Yukawa plasmas near the melting temperature \citep{Caplan2024}. In this work, we argue that diffusion in Yukawa crystals is likely to be normal diffusion or very close to it, rather than anomalous diffusion. 

We briefly review the theory of diffusion as motivation for this work. For a random walker undergoing Brownian motion, a particle diffuses by moving continuously in random directions. In an ensemble of such particles, such as a liquid, the randomness is due to interparticle collisions. When the mean squared displacement (MSD) ${ \langle R(t)^2 \rangle = N^{-1} \sum_{i=1} ^{N} |{\bf r}_i(t+t_0)-{\bf r}_i(t_0)|^2 }$   grows linearly with time we say diffusion is normal, such that ${\langle R(t)^2 \rangle = 6 D t + c}$, with diffusion coefficient $D$ and a small constant offset $c$ to account for ballistic motion at early times before the first collisions. 

However, in the simplest model of diffusion in solids, particles are confined to lattice sites and make a random walk via discrete jumps to nearest neighbor lattice sites with delay times typically taken from a Poisson distribution. In practice, a lattice site must be vacant in order to receive a walker and so diffusion in a perfect crystal would not be observed. Diffusion in real materials therefore requires the thermally activated formation of defects, such as vacancy-interstitial pairs. \cite{Caplan2024} shows that this is the dominant mode of diffusion in the crystallized one-component plasma (OCP) with molecular dynamics simulations. 

Thus, despite mathematical random walks on a lattice being well known to exhibit normal diffusion, diffusion in a true solid is a many-body process. Therefore, diffusion may not necessarily be linear with time but could in general be anomalous such that $\langle R(t)^2 \rangle \propto t^n$, with superdiffusion (subdiffusion) where $n>1$ ($n<1$) \citep[for an introduction see][]{vlahos2008}.

Despite their importance to astrophysics, crystallized plasmas remain poorly studied due to the long simulation times needed to resolve transport. Therefore, we make an effort to verify if the idealized solid OCP does indeed exhibit normal diffusion in this work using a long molecular dynamics simulation.

\section{Methods}

\begin{figure*}[ht!]
\includegraphics[width=0.99\textwidth, trim={125 425 220 10},clip]{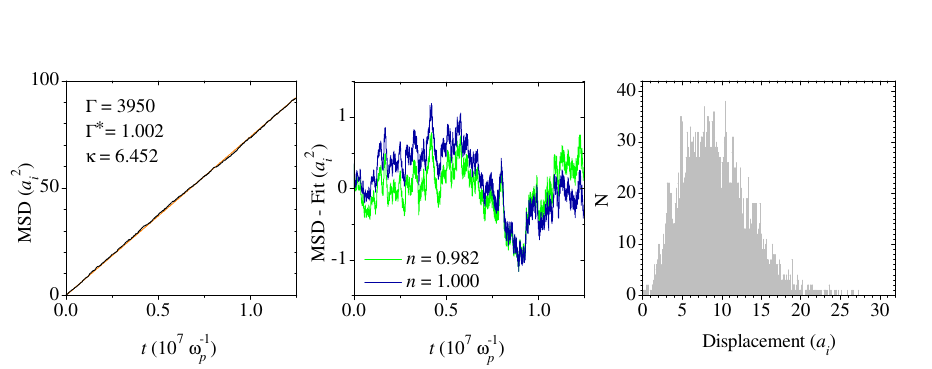}
\caption{Left: The MSD in our simulation (black) grows almost linearly with time; a fit assuming $n=1$ is shown in orange. Center: The residual for our best fit with anomalous diffusion (blue) is not substantially different from the linear fit for normal diffusion (green), and the stochastic variation makes it apparent that the deviation from $n=1$ is just noise. Right: Distribution of particle displacements at the end of the simulation show a characteristic Poissonian envelope, again providing strong evidence for $n=1$.
\label{fig1}}
\end{figure*}

We briefly review our molecular dynamics methods, for a complete description see \cite{Caplan2024}. 
In our OCP, point particles both with charge $eZ$ interact via a screened two-body Coulomb potential

\begin{equation}
    V(r) = \frac{ e^2 Z^2 }{r} e^{-r/\lambda}
\end{equation}

\noindent with interparticle separation $r$ and screening length $\lambda$. For the one-component plasma, the dimensionless plasma coupling factor ${\Gamma = e^2 Z^2/a_i T}$ is defined as a ratio of Coulomb energy to thermal energy, with temperature $T$ and ion sphere radius ${a_i = (4 \pi n_i /3)^{-1/3}}$ from the ion number density $n_i$. 

Together with the dimensionless screening length ${\kappa = n^{-1/3} / \lambda}$, these two dimensionless numbers give the melting temperature of a mixture. While the unscreened OCP crystallizes at ${ \Gamma_{\rm crit} = 175.7}$ \citep{baiko2022ab}, the softening of the potential with higher $\kappa$ pushes the melt line to higher $\Gamma$ (lower $T$). Normalizing to the melting temperature $\Gamma_{\rm M}$ is common. We define a modified coupling parameter ${\Gamma^*  = \Gamma / \Gamma_\mathrm{M}(\kappa)}$ so that the plasma is solid when $\Gamma^* \gtrsim 1$, using ${ \Gamma_\mathrm{M} 
 = \Gamma_{\rm crit} \, e^\kappa (1+\kappa + 0.5 \kappa^2)^{-1}}$ from
\cite{vaulina2000scaling};


We simulate $16 \times 16 \times 16$ bcc unit cells in a periodic cubic box using screening $\kappa = 6.452$ $( a_i / \lambda = 4.0)$ and ${\Gamma = 3950}$  ${(\Gamma^* = 1.002)}$ with a timestep of ${dt = \omega_p^{-1}/382.177}$ (inverse ion plasma frequency  $\omega_p^{-1}$). This timescale is a factor of 4.74 greater relative to the unscreened case using the correction ${\omega_p^{2} = \omega_{p,0}^{2} (1 + \kappa + 0.5 \kappa^2) e^{-\kappa}  }$ with unscreened plasma frequency $~{\omega_{p,0} = (4 \pi e^2 Z^2 n_i / m )^{1/2}}$, such that ${dt = \omega_{p,0}^{-1}/80.47}$. We simulate in the NVT ensemble for $10^9$ timesteps over two weeks on an NVIDIA A100 GPU, a factor of 100 longer than the simulations in \cite{Caplan2024}. 

The high screening factor justifies the small box size, enabling the long run time despite the $\mathcal{O}(N^2)$ interactions. Nearest neighbor forces are $10^2$ stronger than third nearest neighbor forces at this screening, and $10^3$ stronger than fourth nearest neighbor forces.

\section{Results}

In Fig. \ref{fig1} (left) we show the MSD. The stochastic growth of the MSD is due to variations in the small integer number of defects present in the box; the simulation averages about three vacancy-interstitial pairs, but due to the Poisson-like formation and annihilation this varies and it reaches as low as zero when the instantaneous derivative is zero. 

A two-parameter best fit to $\langle R(t)^2 \rangle = D_n t^n$ gives a dimensionless ${D^*_n= 1.63 \times 10^{-6}}$ and $n= 0.982$, with normalization ${D^*_n = D_n / \omega_p^n a^2}$.  This is near exact agreement with Fig. 1 and Eq. 3 in \cite{Caplan2024}. This best fit differs by about $10^{-1} a_i^2$ from the fit assuming normal diffusion where ${D^* = 1.24 \times 10^{-6}}$ ($n=1$) and is indistinguishable on this plot. Including a constant offset $c$ does not change either fit.

In Fig. \ref{fig1} (center) we show residuals of our best fit (green) and a fit assuming $n=1$ (blue). The stochasticity is due to the fluctuating number of defects and no systematic trend that could be evidence of anomalous diffusion is apparent by inspection. Fluctuations in the MSD have magnitude of order one percent the final MSD, comparable to the one percent deviation in the best fit $n$.

The MSD grows to almost $100 a_i^2$, meaning the average particle displaces $10 a_i$ from its initial position. 
A histogram of final particle displacements in Fig. \ref{fig1} (right) also shows that all particles have moved from their initial lattice site with the characteristic Poisson envelope expected of normal diffusion, and the system is well on its way to becoming fully mixed.

\section{Summary}

Our long molecular dynamics simulation of a Yukawa crystal shows no evidence of anomalous diffusion, with a simple fit to $\langle R(t)^2 \rangle = D_n  t^n$ returning $n=1$ within error. In short, we have shown that the results of \cite{Caplan2024} are valid on dynamically long timescales, and can therefore be trusted as a starting point for mesoscopic modeling of neutron star crusts, for example with analytic models of creep or more sophisticated discrete dislocation dynamics simulations. Future work is still needed and several caveats follow. 

We caution against over-interpreting these results, especially for astrophysics. The actual crystal present in neutron star crusts almost certainly has grain boundaries that can act as sources and sinks for defects. Diffusion in these systems may be dominated by the mobility of ions in these thin, disordered interfaces. Strong magnetic fields also permeate the crust, so there is therefore reason to suspect that anomalous diffusion will be found quite naturally in more realistic systems, even if the idealized system simulated in this work is well behaved.

\begin{acknowledgments}
We thank Ellen Zweibel and Robert Ewart. Financial support for this publication comes from Cottrell Scholar Award \#CS-CSA-2023-139 sponsored by Research Corporation for Science Advancement. This work was supported by a grant from the Simons Foundation (MP-SCMPS-00001470) to MC. This research was supported in part by the National Science Foundation under Grant No. NSF PHY-1748958.
\end{acknowledgments}

\software{LAMMPS \citep{LAMMPS}}






\bibliographystyle{aasjournal}



\end{document}